\documentclass[aps,prb,twocolumn,preprintnumbers,amsmath,amssymb,superscriptaddress]{revtex4}%

\usepackage{graphicx}%
\usepackage{dcolumn}
\usepackage{amsmath}
\usepackage{color}

\newcommand{\RN}[1]{\textup{\uppercase\expandafter{\romannumeral#1}}}%

\begin{document}

\title{Superconductivity of Bi-III phase of elemental Bismuth:  \\ insights from Muon-Spin Rotation and Density Functional Theory}

\author{Rustem~Khasanov}
 \email{rustem.khasanov@psi.ch}
 \affiliation{Laboratory for Muon Spin Spectroscopy, Paul Scherrer Institute, CH-5232 Villigen PSI, Switzerland}

\author{Hubertus Luetkens}
 \affiliation{Laboratory for Muon Spin Spectroscopy, Paul Scherrer Institute, CH-5232 Villigen PSI, Switzerland}

\author{Elvezio Morenzoni}
 \affiliation{Laboratory for Muon Spin Spectroscopy, Paul Scherrer Institute, CH-5232 Villigen PSI, Switzerland}

\author{Gediminas Simutis}
 \affiliation{Laboratory for Muon Spin Spectroscopy, Paul Scherrer Institute, CH-5232 Villigen PSI, Switzerland}

\author{Stephan Sch\"{o}necker}
    \affiliation{Applied Materials Physics, Department of Materials Science and Engineering, KTH Royal Institute of Technology, SE-10044 Stockholm, Sweden}

\author{Andreas \"{O}stlin}
    \affiliation{Augsburg Center for Innovative Technologies, and Center for Electronic Correlations and Magnetism, Theoretical Physics III, Institute of Physics, University of Augsburg, D-86135 Augsburg, Germany}

\author{Liviu Chioncel}
    \affiliation{Augsburg Center for Innovative Technologies, and Center for Electronic Correlations and Magnetism, Theoretical Physics III, Institute of Physics, University of Augsburg, D-86135 Augsburg, Germany}

\author{Alex Amato}
 \affiliation{Laboratory for Muon Spin Spectroscopy, Paul Scherrer Institute, CH-5232 Villigen PSI, Switzerland}

\begin{abstract}
Using muon-spin rotation the pressure-induced superconductivity in the Bi-III phase of elemental Bismuth (transition temperature $T_{\rm c}\simeq7.05$~K) was investigated. The Ginzburg-Landau parameter $\kappa=\lambda/\xi=30(6)$ ($\lambda$ is the magnetic penetration depth, $\xi$ is the coherence length) was estimated which is the highest among single element superconductors. The temperature dependence of the superconducting energy gap [$\Delta(T)$] reconstructed from $\lambda^{-2}(T)$ deviates from the weak-coupled BCS prediction. The coupling strength $2\Delta/k_{\rm B}T_{\rm c}\simeq 4.34$ was estimated thus implying that Bi-III stays within the strong coupling regime. The Density Functional Theory calculations suggest that superconductivity in Bi-III could be described within the Eliashberg approach with the characteristic phonon frequency $\omega_{\rm ln}\simeq 5.5$~meV. An alternative pairing mechanism to the electron-phonon coupling involves the possibility of Cooper pairing induced by the Fermi surface nesting.
\end{abstract}


\maketitle


Following the intense research on superconductivity in cuprates and Fe-based superconductors, simpler materials such as elementary or
binary compounds have attracted renewed interest.
Recent results have shown that in spite of their simple composition  they represent a playground for the discovery of
new  phenomena and unconventional characteristics.
Binary compounds were reported to superconduct at surprisingly high critical temperatures ($T_{\rm c}$'s) reaching $\simeq 40$~K for MgB$_2$,\cite{Nagamatsu_Nature_2001} as well as the highest ever obtained $T_{\rm c}$ of $\sim200$~ K in SH$_3$ at the (extreme) pressure of $p\simeq200$~GPa.\cite{Drozdov_Nature_2015} The basic elements with high $T_{\rm c}$'s include Li ($T_{\rm c}~\approx 15-20$~K at 30 GPa), \cite{Struzhkin_Science_2002, Shimizu_Nature_2002, Deemyad_PRL_2003} Ca ($T_{\rm c} \approx 21-29$~K at 220 GPa),\cite{Sakata_PRB_2011} Sc and Y ($T_{\rm c}\approx 20$~K near 100 GPa),\cite{Debessai_PRB_2008, Hamlin_PhysC_2007} V ($T_{\rm c}\approx 17$~K at 120 GPa),\cite{Ishizuka_PRB_2000, Suzuki_JPCM_2002} and S ($T_{\rm c}\approx 17$~K at 220 GPa).\cite{Struzhkin_Nature_1997} Such high transition temperatures of basic elements  indicate the importance of pressure as a tuning parameter of superconducting properties and its role in unraveling the intrinsic properties of the electronic system.

Among single element superconductors the Bi-III phase of elemental Bismuth is one of the most interesting system to study. Following Refs.~\onlinecite{Brandt_JETP_1963, Buckel, Ilina_JETP_1970, Lotter_EPL_1988, Du_PRL_2005, Li_PRB_2017, Brown_ScAdv_2018, Brown_Thesis_2017}, Bismuth converts into a phase Bi-III, exhibiting superconductivity at $T_{\rm c}\simeq 7$~K, above a pressure of $\simeq 2.7$~Gpa. The crystal structure of Bi-III was resolved only recently (in the year 2000) and it was indexed as an incommensurate host-guest lattice.\cite{McMahon_PRL_2000}  It consists of two interpenetrating structures, with ''guest`` atoms forming chains within cylindrical cavities in the ''host`` lattice.
In the $ab-$plane, the unit cells of guest and host match, while 
the ratio along the $c-$axis lattice parameters is incommensurate: $c_{\rm host}/c_{\rm guest} = 1.309$.\cite{McMahon_PRL_2000, Degtyareva_HPR_2005}
%
Remarkably, this incommensurate structure may give rise to 
an additional acoustic mode, in the phonon spectrum, at very low frequencies arising from the sliding of one structure through the other, a process which has almost no energy cost.\cite{Hastings_PRL_1977, Heilmann_PRB_1979, Axe_PRB_1982} Following Refs.~\onlinecite{Brown_Thesis_2017, Brown_ScAdv_2018} in Bi-III such mode was suggested to be responsible for a very high electron-phonon coupling strength and gave rise to the enhanced transition temperature and
the high upper critical field ($H_{\rm c2}$). It is worth to note, however, that the number of physical quantities studied so far for Bi-III phase of elemental Bismuth were mostly limited to $T_{\rm c}$ and $H_{\rm c2}$,\cite{Brandt_JETP_1963, Buckel, Ilina_JETP_1970, Lotter_EPL_1988, Du_PRL_2005,Li_PRB_2017, Brown_ScAdv_2018, Brown_Thesis_2017} which may not be enough to draw unambiguous conclusion on type of the superconducting mechanism.

In this paper, we report on the results of experimental and theoretical studies of Bi-III phase of elemental Bismuth. The measurements of the temperature dependence of the magnetic field penetration depth ($\lambda$) were performed in the muon-spin rotation ($\mu$SR) experiments ($p\simeq 2.72$~GPa). The magnetic field distribution in the sample below $T_{\rm c}$ reflects the formation of a vortex lattice thus confirming that Bi-III is a type-II superconductor.\cite{Li_PRB_2017, Brown_ScAdv_2018, Brown_Thesis_2017} A zero-temperature value of the magnetic penetration depth $\lambda(0)=301(4)$~nm was determined. With the coherence length $\xi=10(2)$~nm taken from
the upper critical measurements,\cite{Li_PRB_2017, Brown_ScAdv_2018} a Ginzburg-Landau parameter $\kappa= 30(6)$ ($\kappa=\lambda/\xi$) was obtained which turns out to be the highest among single element superconductors.
Density Functional Theory (DFT) calculations were used to determine the equilibrium lattice structure parameters of Bi-III under pressure. The DFT results of electronic structure, phonons, and Fermi surfaces, are used to analyze different possible scenarios for the onset
of superconductivity in Bi-III.
Within the framework of the Eliashberg theory, for the given spectral density
of Bi-III, the most effective phonon energy $\omega_{\rm ln}\simeq 5.5$~meV was identified. An alternative pairing mechanism to the electron-phonon coupling which involves the possibility of Cooper pairing induced by the Fermi surface instabilities is also discussed.

\begin{figure*}[htb]
\includegraphics[width=0.95\linewidth,clip]{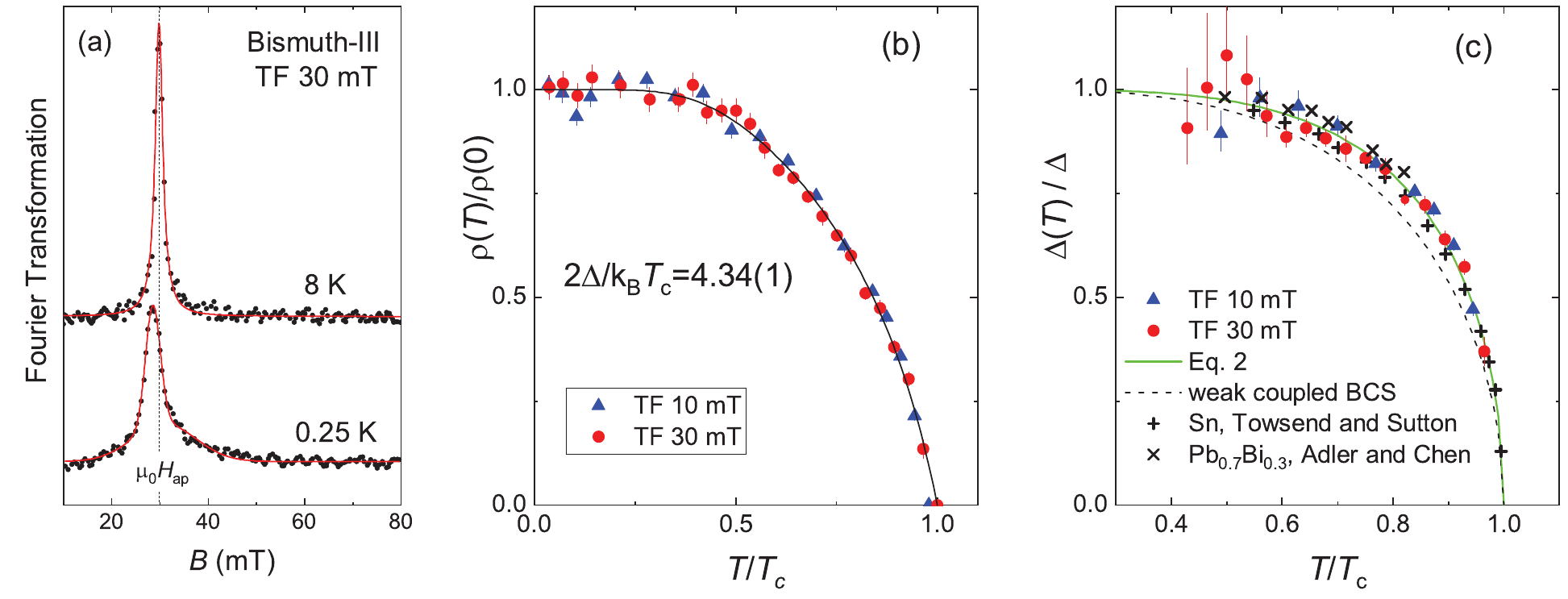}
%
\caption{ (a) Fourier transform of TF-$\mu$SR time spectra ($\mu_0H_{\rm ap}=30$~mT) reflecting the internal field distribution $P(B)$ in the Bi-III sample above ($T=8.0$~K) and below ($T=0.25$~K) the superconducting transition temperature. The stronger broadening and the asymmetric field distribution in the superconducting state is caused by the formation of the vortex lattice. The solid lines are fits (see the Supplemental Part, Ref.~\onlinecite{Supplemntal_part}). (b) The temperature dependence of the superfluid density $\rho(T)/\rho(0)=\lambda^{-2}(T)/\lambda^{-2}(0)$ of Bi-III sample obtained in TF-$\mu$SR experiments at the applied field $\mu_0 H_{\rm ap}=10$~mT (blue triangles) and 30~mT (red circles), respectively. The solid line is the fit of Eq.~\ref{eq:lambda} with the gap described by Eq.~\ref{eq:Gap_nonBCS}. (c) The temperature dependence of the superconducting gap reconstructed from the superfluid density data. The solid line is the fit
of Eq.~\ref{eq:Gap_nonBCS} to the $\Delta(T)$ data. The dashed line is the weak-coupled BCS temperature behavior, Ref.~\onlinecite{Muhlschlegel_ZPB_1959}.
The $+$ and $\times$ symbols refer to $\Delta(T)/\Delta$ points for the superconducting Sn and Pb$_{0.7}$Bi$_{0.3}$ obtained in tunneling experiments
(after Townsend and Sutton, Ref.~\onlinecite{Townsend_PR_1962} and Adler and Chen, Ref.~\onlinecite{Adler_SSC_1971}). }
 \label{fig:Experiment}
\end{figure*}
%
The Bi sample was prepared from commercially available Bi granules (Alfa Aesar, 99.997\% purity). The 'soft' Bi granules were pressed into a cylinder shape ($\simeq5.9$~mm in diameter and $\simeq12$~mm height) and placed inside a double-wall piston-cylinder pressure cell made of MP35/NiCrAl alloy. The construction of the pressure cell is similar to the one described in Refs.~\onlinecite{Khasanov_HPR_2016,Shermadini_HPR_2017}.

The ac susceptibility measurements at $p\simeq 2.72$~GPa reveal the presence of a sharp transition with $T_{\rm c}\simeq 7.05$~K (see the Supplemental Part, Ref.~\onlinecite{Supplemntal_part}). No indication of the transition at $T\sim 3.9$~K was detected, which would correspond to an admixture of the Bi-II phase as is observed, {\it e.g.}, in Ref.~\onlinecite{Li_PRB_2017} for pressures up to $2.9$~GPa.

The transverse-field (TF) $\mu$SR experiments were carried out at the $\mu$E1 beam line by using the dedicated GPD (General Purpose Decay) spectrometer (Paul Scherrer Institute, Switzerland). The details of   TF-$\mu$SR under pressure experiments are provided in the Supplemental Part, Ref.~\onlinecite{Supplemntal_part}, and in the Ref.~\onlinecite{Khasanov_HPR_2016}.
Figure \ref{fig:Experiment}~(a) shows the Fourier transform of TF-$\mu$SR time spectra (the pressure cell background subtracted, see the Supplemental Part, Ref.~\onlinecite{Supplemntal_part}) reflecting the internal field distribution $P(B)$ in the Bi-III sample. The asymmetric $P(B)$ distribution at $T\simeq0.25$~K possesses the basic features expected for an ordered vortex lattice, namely: the cutoff at low fields, the peak shifted below $\mu_0 H_{\rm ap}$ and the long tail towards the high field direction (see {\it e.g.} Refs.~\onlinecite{Maisuradze_JPCM_2009, Khasanov_PRB_2016} and references therein). Our experiments confirm, therefore, that the superconductivity of Bi-III is of a type-II.\cite{Li_PRB_2017, Brown_ScAdv_2018, Brown_Thesis_2017}

\begin{figure*}[htb]
\includegraphics[width=0.9\linewidth,clip]{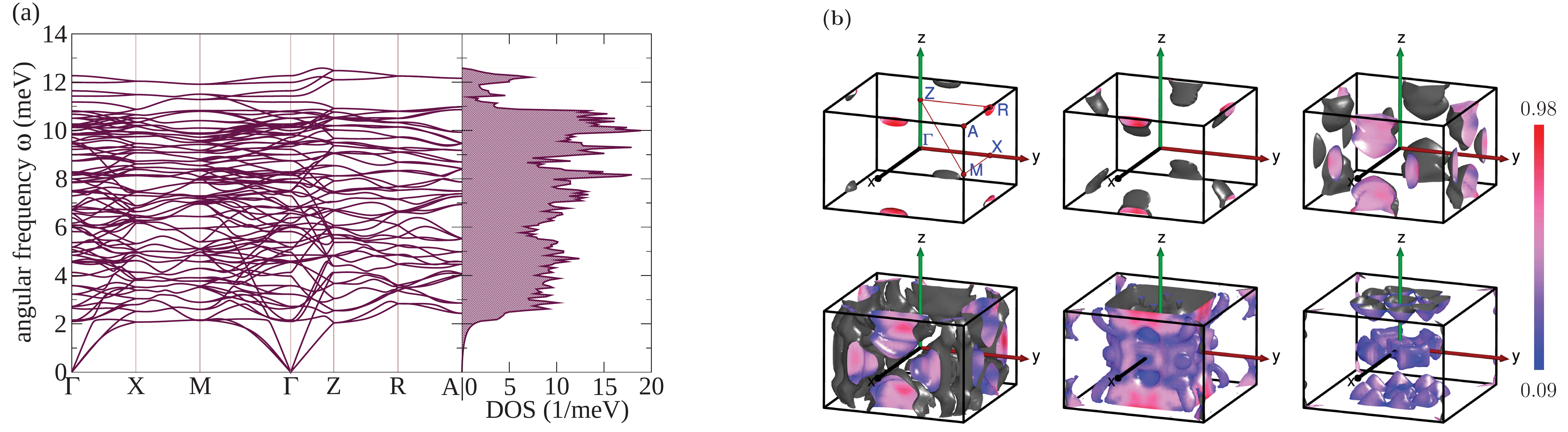}
\caption{(a) Phonon dispersions relations and phonon DOS of the
Bi-\RN{3} approximant at 2.7\,GPa. The logarithmic phonon frequency
$\omega_{\rm ln} \approx 6$\,meV is located in the middle of the phononic
band structure. The DOS is normalized to the number of normal modes per unit cell. (b) Nested multi-sheet Fermi surfaces.
The color code indicates the absolute value of the Fermi velocity in units of $10^6$\,m/s.
High-symmetry points of the Brillouin
zone corresponding to the space group $P4/ncc$
in units of $2\pi(\frac{1}{a}, \frac{1}{a}, \frac{1}{c})$. In calculations the
approximantion of the incommensurate lattice structure of Bi-III by 32 atoms was used.}

 \label{fig:theory}
\end{figure*}

The temperature dependence of the superfluid density [$\rho_s(T)/\rho_s(T=0)=\lambda^{-2}(T)/\lambda^{-2}(T=0)$, Fig.~\ref{fig:Experiment}~(b)] was further obtained from the second central moment of the magnetic field distribution ($\langle \Delta B^2\rangle$).
The superconducting part of the second moment, which for extreme type-II superconductors becomes a direct measure of the
magnetic penetration depth $\lambda$, was calculated via:
$\langle \Delta B^2\rangle_{sc} = \langle \Delta B^2\rangle - \sigma_{\rm nm}^2/\gamma_\mu^2=0.00371 \; {\Phi_0^2}/{\lambda^4}.$\cite{Brandt_PRB_1988, Brandt_PRB_2003}
Here $\sigma_{\rm nm}$ is the contribution of nuclear magnetic moments measured at $T > T_{\rm c}$ and $\Phi_0=2.068\; 10^{-15}$~Wb is the magnetic flux quantum. From the measured $\gamma_\mu\langle \Delta B^2\rangle_{sc}^{1/2}({\rm 0.25~K, 10~mT})=1.22(4)$~$\mu{\rm s}^{-1}$ and $\gamma_\mu\langle \Delta B^2\rangle_{sc}^{1/2}({\rm 0.25~K, 30~mT})=1.15(3)$~$\mu{\rm s}^{-1}$ the value of the magnetic penetration depth was found to be $\lambda(0.25{\rm ~K})=301(4)$~nm. With the coherence length, $\xi= 10(2)$~nm, estimated from the value of the upper critical field
$H_{\rm c2}(T=0)\simeq 2.3-4.6$~T,\cite{Li_PRB_2017, Brown_ScAdv_2018, Brown_Thesis_2017} this translates into a Ginzburg-Landau
parameter $\kappa=\lambda/\xi\simeq 30(6)$. To the best of our knowledge this is the highest $\kappa$ value reported to date for single element superconducting materials. This indicates that Bi-III is a representative of extreme type-II superconductors with $\kappa \gg 1$.

The analysis of the superfluid density data was performed by assuming the superconducting gap of $s-$wave symmetry and using the following functional form: \cite{Tinkham_75}
\begin{equation}
\frac{\rho(T)}{\rho(0)}=  1
+2\int_{\Delta(T)}^{\infty}\left(\frac{\partial
f}{\partial E}\right)\frac{E\
dEd\varphi}{\sqrt{E^2-\Delta(T)^2}}~,
 \label{eq:lambda}
\end{equation}
where $f=[1+\exp(E/k_{\rm B}T)]^{-1}$ is  the Fermi function. The temperature dependence of the superconducting gap $\Delta(T)$ was obtained
by solving the nonlinear Eq.~\ref{eq:lambda} for every temperature point.
The resulting $\Delta(T)/\Delta$ values for Bi-III are presented in Figure~\ref{fig:Experiment}~(c). Here $\Delta$ is the zero-temperature value of the
superconducting energy gap. Note that due to saturation of the superfluid density data [$\rho(T)/\rho(0)\simeq 1$ for $T/T_{\rm c}\lesssim 0.4$, Fig.~\ref{fig:Experiment}~(b)] the low-temperature $\Delta(T)$ values can not be obtained with reliable accuracy.
Figure~\ref{fig:Experiment}~(c) implies that upon decreasing temperature, the gap in Bi-III grows faster than  expected for the weak-coupled BCS $\Delta(T)$.\cite{Muhlschlegel_ZPB_1959} For comparison, the results for the superconducting gap in Sn, Ref.~\onlinecite{Townsend_PR_1962}, and Pb$_{0.7}$Bi$_{0.3}$, Ref.~\onlinecite{Adler_SSC_1971},
which were found to be characterized by non-BCS $\Delta(T)$, are also shown in Fig.~\ref{fig:Experiment}~(c). Remarkably, the
temperature dependence of the superconducting gap in Sn, Pb$_{0.7}$Bi$_{0.3}$ and Bi-III  have a very similar functional form.

The temperature dependence of the gap presented in Fig.~\ref{fig:Experiment}~(c)
was approximated by the equation:\cite{Prozorov_SST_2008}
\begin{equation}
\Delta(T)=\Delta\;\tanh \left[ \frac{\pi k_{\rm B}}{\Delta}\sqrt{c \left( \frac
{T_{\rm c}}{T} -1   \right) }  \right]
 \label{eq:Gap_nonBCS}
\end{equation}
with $c$ and $\Delta$ as fit parameters.  Following Ref.~\onlinecite{Prozorov_SST_2008} this general equation describes $\Delta(T)$'s for superconductors with
various coupling strengths ($2\Delta/k_{\rm B}T_{\rm c}$) and order parameter symmetries. The fit of Eq.~\ref{eq:Gap_nonBCS}
to the data with $c=2.31(2)$ and $\Delta=1.32(1)$~meV is presented in Fig.~\ref{fig:Experiment}~(c) by a solid green line.
The solid line in Fig.~\ref{fig:Experiment}~(b) corresponds to the fit of Eq.~\ref{eq:lambda} with such obtained $\Delta(T)$ to the superfluid density data.
Using the value of $2\Delta/k_{\rm B}T_{\rm c}\simeq 4.34$ in Carbotte's empirical relation:\cite{Carbotte_RMP_1990}
\begin{equation}
\frac{2 \Delta}{k_{\rm B} T_{\rm c}}=3.53 \left[ 1+12.5 \left( \frac{k_{\rm B}T_{\rm c}}{ \omega_{\rm ln}}\right)^2  \ln \frac{\omega_{\rm ln}}{2 k_{\rm B}T_{\rm c}} \right],
 \label{eq:Gap_Carbotte}
\end{equation}
the logarithmically averaged phonon frequency $\omega_{\rm ln}\simeq 5.51$~meV was calculated. In fact, $\omega_{\rm ln}$
corresponds to the dynamics of the superconducting state
and represents the most effective phonon energy for a given $T_{\rm c}$. Following phenomenological arguments of Carbotte,\cite{co.ca.88,Carbotte_RMP_1990} for Bi-III one expects $\omega_{\rm ln}\propto 7 k_{\rm B}T_{\rm c}\simeq 4.3$~meV which is close to 5.51~mev estimated from Eq.~\ref{eq:Gap_Carbotte}.

It is worth to emphasize that even though the temperature dependence of the gap in most of the cases was found to follow the weak-coupled BCS prediction, a non-BCS gap behavior is not something unexpected. Prominent examples are the metallic Sn  and Pb,\cite{Gasparovic_SSC_1966} in the case of single element superconductors, and MgB$_2$\cite{Lee_PhyC_2002, Ponomarev_SSC_2004, Brinkman_PRB_2002} and PbBi alloys\cite{Adler_SSC_1971} for binary ones. The deviation from weak-coupled BCS $\Delta(T)$ is quite often observed in cuprate and Fe-based high-temperature superconductors. Theoretically, a non-BCS behavior of $\Delta(T)$ is expected for superconductors with high values of the electron-phonon coupling constant, within the Eliashberg-Nambu formalism,\cite{Leavens_PRB_1984, Zheng_PRB_2008} and is explained by damping of quasiparticle excitations caused by a strong electron-phonon interaction.\cite{Schrieffer_BAS_1963, Wada_RMP_1964, Wada_PR_1964}


In the following part of the paper the results of Density
Functional Theory calculations are discussed. The structural analysis and the electronic band
structure are presented in the Supplementary Part, Ref.~\onlinecite{Supplemntal_part}.
The calculations reveal that at ambient pressure, where Bi adopts a
rhombohedral structure (Bi-I phase), its density of states (DOS) shows bands which are clearly
separated into bonding/antibonding $s-$ and $p-$states. The Fermi level, $E_F$, is
situated in a gap between the bonding and antibonding $p$-states thus giving rise to the semimetallic behavior of Bi-I.\cite{Hausermann_JACS_2002}
As pressure increases, the $p$-states start to overlap, leading to a metallic
DOS at $E_F$.
From the analysis of electronic bands and DOS (see the Supplementary Part, Ref.~\onlinecite{Supplemntal_part}) the host and guest
sites in Bi-III phase were found to have similar magnitudes, in agreement with Ref.~\onlinecite{Hausermann_JACS_2002}.
The electronic bands of $p$-character would provide electrons for the pairing.

Figure~\ref{fig:theory}~(a) shows the phonon dispersion and the phonon DOS obtained with the approximation of the incommensurate lattice structure of Bi-III by 32 atoms. A different structure of 42 atoms was considered in Ref.~\onlinecite{Brown_ScAdv_2018} and it was found to demonstrate the presence of a low frequency phason mode. This mode was suggested to be responsible for a very high electron-phonon coupling strength and gave rise to the enhanced transition temperature and the high upper critical field.\cite{Brown_ScAdv_2018,Brown_Thesis_2017}  Note, however, that in the actual calculations with 32 atoms {\it no phason modes} were found to appear in the phonon spectra. One would also mention that in accordance with arguments of the Eliashberg theory, neither the very low nor the very high frequencies are important, while frequencies around the middle of the spectrum, $\approx 6$~meV [see Fig.~\ref{fig:theory}~(a)], are responsible for the magnitude of $T_{\rm c}$.\cite{Carbotte_RMP_1990} Note that this value stays close to $\omega_{\rm ln}\simeq 5.51$~meV obtained from measured
$2\Delta/k_{\rm} T_{\rm c}\simeq 4.34$ by means of Eq.~\ref{eq:Gap_Carbotte}. The Eliashberg theory provides no arguments on
limiting $T_{\rm c}$ for a given pairing mechanism (phonon, phason or any other boson exchange).
Considering the electron-phonon interaction, the limitation on the magnitude of $T_{\rm c}$ for Bi-III is most probably related to the lattice (in)stability for a given value of the applied pressure.

An alternative superconducting pairing mechanism,
based on correlations in the Fermi liquid
rather than electron-phonon coupling, was proposed by
Kohn and Luttinger.\cite{ko.lu.65}
This pairing mechanism is associated with the effective
interaction,
between quasiparticles occurring
as a result of polarization of the fermionic background,
which is proportional to the static susceptibility (Lindhard function).
The appearance of a divergence in the static susceptibility
$\chi({\bf q})$
is determined by the existence of nesting vectors ${\bf q}_{nest}$
arising if the Fermi surface
possesses parallel fragments such that pairs of electronic
states can be connected by the same wave vector ${\bf q}_{nest}$.
In this case the possibility of Cooper pairing is determined by
the characteristics of the energy spectrum (band structure) in
the vicinity of the Fermi level and by the effective interaction.
In Fig.~\ref{fig:theory}~(b) the results of the Fermi surface (FS) calculations are shown.
Based on the FS shape one can not exclude that such a simple mechanism
of Fermi surface instabilities underlies the
pairing formation in Bi-III, in particular in conjunction with the
Kohn-Luttinger mechanism.\cite{ko.lu.65} Moreover, due to
proximity to a maximum in the DOS, the
superconducting transition temperature
(predicted within the framework Kohn-Luttinger mechanics)
may increase further
as a consequence of the combination with van Hove
singularities.\cite{hlub.99}

To conclude, muon-spin rotation experiments performed under pressure $p\simeq2.72$~GPa indicate that the Bi-III phase of the elemental Bismuth is a type-II superconductor characterized by a large Ginzburg-Landau parameter  $\kappa= 30(6)$. The coupling strength
$2\Delta/k_{\rm B}T_{\rm c}\simeq 4.34$ was found thus implying that Bi-III lies in the strong coupling regime. The temperature behavior of the
superfluid density was found to be consistent with a single gap of $s-$wave symmetry with $\Delta=1.32(1)$~meV.
Based on DFT results for the electronic and
phononic band structures and Fermi surface,
a possible mechanism of superconducting pairing in Bi-III was qualitatively discussed.
The DFT results, in combination with the
Eliashberg's theory arguments, reveal that phonons
with frequency around $5.5$~meV are most effective in
producing $T_{\rm c} \approx 7$K.
An alternative pairing mechanism to the electron-phonon coupling involves the possibility of Cooper pairing induced by the Fermi surface instabilities. FS calculations indicate the existence of nesting vectors, so the pairing mechanism of Kohn-Luttinger type may also be considered.

This work was performed at the Swiss Muon Source (S$\mu$S), Paul Scherrer Institute (PSI, Switzerland). The work of GS is supported by the Swiss National Science Foundation, grants $200021 \_ 149486$ and $200021 \_ 175935$.
L.C. and A.\"O acknowledges DFG for the financial support through TRR80/F6
project.
S.S. acknowledges financial support by the Swedish Research Council and computational facilities provided by the Swedish National Infrastructure for Computing at the National Supercomputer Centers in Link\"oping and Ume\aa. RK is grateful to E. Pomyakushina for providing granular Bismuth.

\end{document}